\documentclass[prd,aps,twocolumn,amsfonts,showpacs,superscriptaddress,preprintnumbers,nofootinbib]{revtex4-2}

\usepackage{graphicx}
\usepackage{xcolor}
\usepackage{rotating}
\usepackage{bm,amsmath,amssymb}
\usepackage[utf8]{inputenc}
\usepackage{comment}
\usepackage{footmisc}
\usepackage{diagbox}
\usepackage{amsmath}
\usepackage{cancel}
\usepackage{physics}
\usepackage{dcolumn}% Align table columns on decimal point
\usepackage{bm}% bold math

\usepackage[colorlinks=true, urlcolor=blue, linkcolor=blue, citecolor=blue, hyperindex=true, linktocpage=true]{hyperref}

\newcommand{\sh}[1]{\sinh{\frac{\beta \omega_{#1}}{2}}}

\begin{document}

%\preprint{}

\title{Duality Constraints on Thermal Spectra of 3$d$ Conformal Field Theories and 4$d$ Quasinormal Modes}

\author{Sa\v{s}o Grozdanov}
\affiliation{Higgs Centre for Theoretical Physics, University of Edinburgh, Edinburgh, EH8 9YL, Scotland, 
}
\affiliation{Faculty of Mathematics and Physics, University of Ljubljana, Jadranska ulica 19, SI-1000 Ljubljana, Slovenia
}

\author{Mile Vrbica}
\affiliation{Higgs Centre for Theoretical Physics, University of Edinburgh, Edinburgh, EH8 9YL, Scotland,
}

\begin{abstract}
Thermal spectra of correlation functions in holographic 3$d$ 
large-$N$ conformal field theories (CFTs) correspond to quasinormal modes of classical gravity and other fields in asymptotically anti-de Sitter black hole spacetimes. Using general properties of such spectra along with constraints imposed by the S-duality (or the particle-vortex duality), we derive a spectral duality relation that all such spectra must obey. Its form is universal and relates infinite products over QNMs with bulk algebraically special frequencies. In the process, we also derive a new sum rule constraining products over QNMs. The spectral duality relation, which imposes an infinite set of constraints on the QNMs, is then investigated and a number of well-known holographic examples that demonstrate its validity are examined. Our results also allow us to understand several new aspects of the pole-skipping phenomenon. 
\end{abstract}

\maketitle

%\tableofcontents

\section{Introduction}

Three-dimensional conformal field theories (CFTs) are an immensely rich family of quantum field theories with applications ranging from condensed matter systems at critical points to high-energy physics, and via the holographic duality, to four-dimensional classical and quantum gravity (see e.g.~Refs.~\cite{sachdev_2011,zaanen2015holographic,hartnoll2018holographic}). 3$d$ CFTs are special in that they exhibit the S-duality associated either to the $U(1)$ current (known as the particle-vortex duality) \cite{witten2003sl2z} or the energy-momentum tensor \cite{Leigh:2003ez}. In terms of the gravitational bulk, this amounts to exchanging the electric and magnetic components of gauge fields, which, in a number of systems, imposes self-duality relations on the two-point correlators.

In this work, we generalise such duality relations to \emph{thermal} two-point correlators of composite operators in holographic large-$N$ CFTs and derive a universal relation relating their spectra. The prototypical example of such a theory is the theory that describes a large-$N$ stack of M2 branes \cite{Maldacena:1997re,ABJM,Herzog:2002fn,Herzog:2007ij}. Since holography establishes that such spectra equal those of quasinormal modes (QNMs) in the 4$d$ bulk \cite{Kovtun:2005ev}, our results therefore independently apply to the linearised fluctuations of classical fields (gravity, Maxwell fields, etc.) in the backgrounds of asymptotically anti-de Sitter (AdS) black holes. 

In 3$d$ CFTs or 4$d$ gravity, correlators or bulk perturbations can be organised into longitudinal (or even, denoted by $+$) and transverse (or odd, denoted by $-$) channels with respect to the spatial wavevector $\vb{k}$ direction. Retarded correlators $G_\pm(\omega,{\bf{k}})$ then control real-time response and can be analytically continued to give other two-point functions. Dualities relating pairs of bulk equations of motion in a large class of theories \cite{chandrasekharMathematicalTheoryBlack1983,lenziDarbouxCovarianceHidden2021,Grozdanov:2023txs,bakasEnergymomentumCottonTensor2009,glampedakisDarbouxTransformationBlack2017} imply a {\em duality}-type relation between the CFT correlators in the two channels:
\begin{equation}
     G_+(\omega,k) G_-(\omega,k)=\frac{\omega^2}{\omega_*^2(k)}-1, \label{eq:duality}
\end{equation}
where $\omega$ is the frequency and $k \equiv |\bf{k}|$. From the point of view of the bulk, $\omega_*$ are the 
{\em algebraically special frequencies} (see Refs.~\cite{bakasEnergymomentumCottonTensor2009,Bakas_2014}). As described in \cite{Grozdanov:2023txs}, they are related to {\em pole skipping} \cite{Grozdanov:2017ajz,Blake:2017ris,Blake:2018leo,Grozdanov:2018kkt}. For all correlators of interest, $G(\omega,k)$ are meromorphic functions in $\omega$, with each $G_+$ and $G_-$ containing an infinite set of poles (or QNMs) in the spectra with dispersion relations $\omega^+_n(k)$ and $\omega^-_n(k)$, respectively. Using the duality \eqref{eq:duality} and general analytic properties of $G(\omega,k)$ in 3$d$ CFTs with holographic duals, in particular, the recent {\em thermal product formula} of Ref.~\cite{Dodelson:2023vrw}, we derive a {\em universal} relation for the QNMs combining the spectral information of both channels. For the following (infinite convergent) product $S(\omega,k)$, defined as 
\begin{equation}
    S(\omega,k) \equiv \qty(1+\frac{\omega}{\omega_*(k)})\prod_n \qty[1-\frac{\omega}{\omega_n^+(k)}]\qty[1+\frac{\omega}{\omega_n^-(k)}], \label{def:S}
\end{equation} 
we show that the odd-in-$\omega$ part of $S(\omega,k)$ is fixed to
\begin{equation}
    S(\omega,k) - S(-\omega,k) =2 i \lambda(k) \sh{}, \label{eq:mainS}
\end{equation}
where the function $\lambda(k)$ can itself be expressed in terms of $\omega_*$ and $\omega^\pm_n$ (see below), and $\beta = 1/T$ is the inverse temperature. Eq.~\eqref{eq:mainS}, which we refer to as the \emph{spectral duality relation}, depends on the state- and correlator-dependent algebraically special frequencies $\omega_*$, which are easily computable from the 4$d$ bulk dynamics. It imposes remarkably stringent, infinite sets of constraints on the spectra of certain 3$d$ CFTs and, thereby, on 4$d$ classical field theories such as General Relativity and the Maxwell theory.

From the point of view of holography, our work extends past studies \cite{Herzog:2007ij,Davison_2015,Murugan:2014sfa,Alejo:2019hnb} that utilised the form of the duality \eqref{eq:duality} in special cases: namely, when, as a result of tuning some parameter of the theory, $\omega_* \to \infty$. Then, Eq.~\eqref{eq:duality} gives what we refer to as the {\em self-duality} relation:
\begin{equation}
     G_+(\omega,k) G_-(\omega,k)= -1. \label{eq:self_duality}
\end{equation}
In the boundary CFT, at this special point, the operator that enters $G_\pm$ (e.g., $T^{\mu\nu}$ or $J^{\mu}$) has a vanishing expectation value. Furthermore, this paper also demonstrates a precise way in which a subset of all pole-skipping points contained in $\omega_*$ can impose an infinite set of constraints on spectra, thereby extending the ideas of Ref.~\cite{Grozdanov:2023tag}. 

We begin by deriving Eq.~\eqref{eq:mainS}. Then, we explore its general implications and end by demonstrating its validity in a few well-known holographic examples.   

\section{Derivation of the spectral duality relation}

Consider retarded finite-temperature correlators $G_\pm(\omega)$ of composite (typically, conserved) operators, such as the energy-momentum tensor $T^{\mu\nu}$ and currents $J^{\mu}$. Hereon, we will suppress any explicit dependence on $k$. In strongly coupled, large-$N$ CFTs with classical holographic duals at $T\neq 0$, all such correlators are meromorphic functions with only simple poles and no branch cuts in the complex $\omega$ plane (see~Refs.~\cite{Gulotta_2011,Grozdanov:2016vgg}). For $k\in\mathbb{R}$, retarded $G_\pm (\omega)$ are holomorphic in the upper half-plane ($\Im\omega > 0$) and $G_\pm(\omega)^*=G_\pm(-\omega^*)$, which implies that the  $\omega_n$ are either imaginary or come in pairs of $\omega_n=-\omega_{n'}^*$. Moreover, the large-$\omega$ asymptotics of $G_\pm(\omega)$ are assumed to be given by 
\begin{equation}
        \omega \rightarrow \infty: \quad G_\pm(\omega)\sim \omega^{m_\pm}, \label{eq:Gasy}
\end{equation}
for some integers $m_\pm$, provided that the limit is taken in a direction in the complex $\omega$ plane that avoids all the poles (see Ref.~\cite{Gulotta_2011}). The presence of logarithmic terms in Eq.~\eqref{eq:Gasy} is ruled out by subsequent analysis.

From the point of view of the 4$d$ bulk, we consider theories where there exist first-order differential operators that map the even and the odd solutions to each other for all $\omega^2 \neq \omega^2_*$, where $\omega_*$ is the algebraically special frequency with $\Re \omega_*=0$ for real $k$. The bulk structure of the duality then gives the following constraint: $G_+(\omega_*)=G_-(-\omega_*)=0$. The QNM analysis also reveals the behaviour of large-$\omega$ asymptotics of $\omega_n^\pm$ \cite{Cardoso_2004,natario2005classification,Motl:2003cd,Dodelson:2023vrw}. In particular, we assume that the poles of $G_\pm(\omega)$ are organised into asymptotic lines indexed by $j$:
\begin{equation}
    \omega_{n,j}=-r_j n e^{i\theta_j}-s_j e^{i\phi_j}+\mathcal{O}(n^{-1/2}), \label{eq:asymptotics}
\end{equation}
for $r_j, s_j >0$ (this includes the standard `Christmas tree' configuration). The lines are either imaginary ($\theta_j=\phi_j=\pi/2$) or the line $j$ has a corresponding mirror image $j'$ ($\omega_{n,j}=-\omega^*_{n',j'}$). Furthermore, we assume
\begin{equation}
    \omega_{n,j}^+-\omega_{n,j}^-=\sigma_j r_j e^{i\theta_j}+\mathcal{O}(n^{-1/2}), \label{eq:offset}
\end{equation}
which means that the asymptotic lines of poles are the same in both channels, but with relative offsets between the modes $\sigma_{j}$. We also introduce the \emph{offset index}: $\sigma = \sum_j \sigma_j$. An indexing scheme should be chosen so that $n \in \mathbb{Z}_+$ for all branches $j$, with every pole counted exactly once. In practice, this means that $\sigma$ counts which channel has `more' poles than the other: the addition of a pole in the longitudinal (transverse) channel increases (decreases) $\sigma$ by one.

The last ingredient needed to derive Eq.~\eqref{eq:mainS} is the thermal product formula of Ref.~\cite{Dodelson:2023vrw}. It can be expressed as 
\begin{equation}\label{eq:TPF}
G_\pm(\omega)-G_\pm(-\omega)  = \frac{2i\lambda_\pm \sh{}}{\prod_n \qty[1-\qty(\frac{\omega}{\omega_n^\pm})^2]}, 
\end{equation}
where $\lambda_\pm$ do not depend on $\omega$. Due to spectral positivity,
\begin{equation}
  \lambda_\pm=-\frac{2i}{\beta}\partial_\omega G_\pm(\omega) |_{\omega=0}>0. \label{def:lambda}
\end{equation}
Eq.~\eqref{eq:TPF} therefore states that the symmetrised Wightman correlator is completely determined, up to a function of $k$, by its poles. This is expected to hold for all holographic correlators that admit a (classical) bulk wave equation description (see Ref.~\cite{Dodelson:2023vrw}). In fact,  Eq.~\eqref{eq:TPF} itself provides constraints on the QNMs in the absence of any discussion of dualities \eqref{eq:duality} that are specific to 3$d$ theories. We express this fact through the following sum rule, which holds for all $k$, and is derived in Appendix~\ref{app:sum}:
\begin{equation}
    \int_\mathbb{R}\frac{d\omega}{\pi}\frac{ \sinh{\frac{\beta\omega}{2}}}{(\omega-i0^+)^2 \prod_n \qty[1-\qty(\frac{\omega}{\omega_n})^2]}=\frac{i\beta}{2}. \label{eq:sumrule}
\end{equation}
The expression is valid provided that the integral converges. The sum rule \eqref{eq:sumrule} holds independently in each of the channels, for all $k$, and in any number of dimensions, thereby complementing those developed in Ref.~\cite{Dodelson:2023vrw}.

Assumptions \eqref{eq:asymptotics} and \eqref{eq:offset} establish the large-$\omega$ asymptotics of the following infinite product (away from the lines of poles):
\begin{equation}
    \omega \rightarrow \infty: \quad \prod_n \frac{1-\frac{\omega}{\omega_n^+}}{1-\frac{\omega}{\omega_n^-}}\sim \omega^{\sigma}. \label{eq:large_w}
\end{equation}
The above-stated properties of $G_\pm(\omega)$ and the duality relation \eqref{eq:duality} then allow us to write the following mixed-channel product representations of $G_\pm$:
\begin{align}
    G_+(\omega)&=\alpha \qty(\frac{\omega}{\omega_*}-1) \prod_n \frac{1-\frac{\omega}{\omega_n^-}} {1-\frac{\omega}{\omega_n^+}},\label{eq:Gpm-1}  \\
    G_-(\omega)&=\frac{1}{\alpha} \qty(\frac{\omega}{\omega_*}+1) \prod_n \frac{1-\frac{\omega}{\omega_n^+} }{1-\frac{\omega}{\omega_n^-}}  .\label{eq:Gpm-2}
\end{align}
Crucially, the assumption \eqref{eq:Gasy} and the large-$\omega$ behaviour of the infinite product $\eqref{eq:large_w}$ dictate that $\alpha$ cannot be a function of $\omega$. This follows from the fact that, to preserve the correct analytic structure of $G_\pm$, $\alpha$ and $1/\alpha$ must be holomorphic in the entire complex plane and (due to Eqs~\eqref{eq:Gasy} and \eqref{eq:large_w}) asymptotically bounded by a polynomial. A standard theorem in complex analysis whereby an entire (everywhere holomorphic) function that is bounded by a polynomial must itself be a polynomial then implies that $\alpha$ is constant in $\omega$.

Eqs.~\eqref{eq:Gasy} and \eqref{eq:large_w} then imply that $\sigma\in \mathbb{Z}$ and Eqs.~\eqref{eq:Gpm-1}--\eqref{eq:large_w} imply that for the self-duality relation \eqref{eq:self_duality}, $m_\pm=\mp \sigma$, and for the duality relation \eqref{eq:duality}, $m_\pm=\mp \sigma + 1$. While we are not aware of a simple method to compute $\sigma$, given that it characterises the low-$\omega$ parts of the spectra, its value is usually clear from a numerical computation of a few lowest-lying QNMs. We stress that, therefore, the low-$\omega$ properties of the spectrum determine the large-$\omega$ scaling and spectral properties of correlators through $\sigma$.

Finally, the product representations \eqref{eq:Gpm-1}--\eqref{eq:Gpm-2} and the thermal product formula \eqref{eq:TPF} lead to our main result: the universal spectral duality relation in Eq.~\eqref{eq:mainS}. The result itself allows us to express $\lambda(k)$ as
\begin{equation}\label{eq:lambda}
\lambda(k) = \frac{2}{i\beta} \left[\frac{1}{\omega_*(k)} + \sum_{n} \left( \frac{1}{\omega^-_n(k)} - \frac{1}{\omega^+_n(k)} \right)  \right].
\end{equation}
The remaining functions of $k$ that appear above are related as 
$\lambda^2 = \lambda_+\lambda_-$ and $\alpha^2=\lambda_+/\lambda_-$, with $\lambda/\alpha >0$. Importantly, $S(\omega)$ defined in \eqref{def:S} encodes all information about the pole structure of both even and odd channels in the sense that the zeros of $S(\omega)/(\omega+\omega_*)$ are the poles of $G_+(\omega)G_-(-\omega)$. Most remarkably, the product is completely determined by a single function $\omega_*(k)$.

\section{General properties and implications of the spectral duality relation}

The spectral duality relation \eqref{eq:mainS} reveals a number of general properties and constraints that the QNM spectra must obey. For example, it implies that if $\omega_*\neq 0$, the poles in the two channels cannot coincide. Interestingly, in all known cases, $\omega_*(k=0)=0$, and therefore $\omega_*$ acts as a `hydrodynamic' gapless mode in the product $S(\omega)$. This also implies that the limits of $k\rightarrow 0$ and $\omega_*\rightarrow \infty$ do not commute. It is reasonable to expect that this is a general feature of the self-dual limit \eqref{eq:self_duality}. Moreover, Eq.~\eqref{eq:mainS} makes it easy to compute $S(\omega_n^-)$, $S(-\omega_n^+)$ or $S(\omega_*)$ as $S(-\omega_n^-)=S(\omega_n^+)=S(-\omega_*)=0$.

Eq.~\eqref{eq:mainS} imposes an infinite number of constraints on an infinite number of QNMs $\omega^\pm_n$. One way to express them is through the introduction of elementary symmetric polynomials $e_i(\mathcal{W})$ evaluated on a truncated set of QNMs: $\mathcal{W} = \{1/\omega_*, 1/\omega^-_1,\ldots,1/\omega^-_n, - 1/\omega^+_1,  \ldots, -1/\omega^+_n \}$, which contains $2n+1$ elements. It is then easy to show that for all $j \in \{0,1,\ldots,n\}$ (we think of taking $n\to\infty$),
\begin{equation}
    e_{2j+1} (\mathcal{W}) = \frac{i\lambda}{(2j+1)!} \left( \frac{\beta}{2}\right)^{2j+1}.
\end{equation}
Eq.~\eqref{eq:lambda} for $\lambda$ follows from the $j=0$ case. Another instructive way to state the constraints is by evaluating Eq.~\eqref{eq:mainS} at Matsubara frequencies $\Omega_\ell=2\pi i \ell/\beta$, which yields a relation between even and odd QNMs:
\begin{equation}
    \qty(1+\frac{\Omega_\ell}{\omega_*})\prod_n \frac{1-\frac{\Omega_\ell}{\omega_n^+}}{1+\frac{\Omega_\ell}{\omega_n^+}}=\qty(1-\frac{\Omega_\ell}{\omega_*})\prod_n \frac{1-\frac{\Omega_\ell}{\omega_n^-}}{1+\frac{\Omega_\ell}{\omega_n^-}},
        \label{eq:matsubara0}
\end{equation}
where $\ell \in \mathbb{Z}$. What is remarkable is that Eq.~\eqref{eq:matsubara0} holds for \emph{any} $\ell$. The equation can therefore again be seen as providing an infinite number of (non-linear) constraints on the QNMs in two separated channels. These identities, for example, allow us to express any $\omega^\pm_n$, as well as $\omega_*$, in terms of all other QNMs in an infinite number of ways.

Finally, we can also use Eq.~\eqref{eq:mainS} to elucidate the fact that in CFTs obeying the duality relation \eqref{eq:duality}, the infinite set of pole-skipping points where retarded $G(\omega)$ exhibit crossing lines of zeros and poles in the $(\omega,k)$ space (a `$0/0$') can only exist at the Matsubara frequencies $\Omega_\ell$ (see also Refs.~\cite{Grozdanov:2017ajz,Blake:2017ris,Blake:2018leo,Grozdanov:2018kkt,Grozdanov:2019uhi,Blake:2019otz} and \cite{Natsuume:2019xcy,Ahn:2020baf,Wang:2022mcq,Grozdanov:2023txs}). Denote a pole-skipping point in each of the channels by $\tilde\omega^\pm$. Eq.~\eqref{eq:duality} then implies that either $\tilde\omega_\pm^2=\omega_*^2$ or that the pole-skipping point is shared between the channels so that  $\tilde\omega_+=\tilde\omega_-$ at some $\tilde k_+=\tilde k_-=\tilde k$ \cite{Grozdanov:2023txs}. It follows that pole skipping can occur if and only if $S(\tilde\omega,\tilde k)=S(-\tilde\omega,\tilde k)=0$. The spectral duality relation \eqref{eq:mainS} then implies that, as claimed, this is only possible if $\tilde\omega=\Omega_\ell$.

\section{Holographic examples}

Next, we place a number of extensively studied holographic examples in the context of our spectral duality relation \eqref{eq:mainS}. Beyond a new demonstration of how stringently constrained such spectra really are and the exploration of the relation's implications, we also use some of these examples to explicitly verify the statements and assumptions used in the derivation of the relation.

First, consider the AdS${}_4$-Schwarzschild black brane. It provides a bulk background that allows for calculations of (decoupled) $T^{\mu\nu}$ and $J^\mu$ correlators $G_\pm^T$ and $G_\pm^J$ in a neutral thermal state. The retarded $G_\pm^J$ correlators obey the self-dual relation \eqref{eq:self_duality}, which was analysed in relation to transport in Ref.~\cite{Herzog:2007ij}. By choosing $\bf{k}$ to point in the $z$-direction (of the $(t,y,z)$ boundary coordinates), we can express the even and the odd correlators as
\begin{align}\label{def:JJ}
    G^J_+=\frac{1}{\gamma k^2}\frac{\delta \expval{J^t}}{\delta A_t} ,\quad
    G^J_-=\frac{1}{\gamma}\frac{\delta \expval{J^y}}{\delta A_y} ,
\end{align}
where $\gamma$ is a normalisation constant. Only the spectrum of $G^J_+$ contains a gapless hydrodynamic (charge diffusion) mode $\omega = - i  D_c k^2 + \mathcal{O}(k^4)$. The offset index is $\sigma=1$, which can be checked numerically (with $\sigma_1 = \sigma_2 = 1/2$). We now expand Eq.~\eqref{eq:mainS} for small $k$ and use the fact that due to isotropy at $k=0$, the gapped parts of the spectra of $G^J_\pm$ coincide. Then,
$\lambda\sim -2/D_c\beta k^2+\mathcal{O}(k^{-1})$ and  
\begin{equation}
\frac{i D_c \beta}{2}\lim_{k\rightarrow 0} k^2 S(\omega)=\sinh\frac{\beta \omega}{2},
\end{equation}
which, interestingly, fixes the spectrum of gapped `poles' at $k=0$ to be the lower half-plane Matsubara frequencies, albeit with zero residues. As $k$ is decreased, this formation of poles along imaginary $\omega$ is reached from the usual QNM `Christmas tree' at large $k$ in a `zipper-like' manner through a sequence of pair-wise pole collisions (see Refs.~\cite{Witczak-Krempa:2012qgh,Witczak_Krempa_2013}). As a result of the cancellations in Eqs.~\eqref{eq:Gpm-1} and \eqref{eq:Gpm-2}, at $k=0$, $G^J_\pm$ are completely determined by the hydrodynamic mode, $G_\pm(\omega,k=0)\propto \omega^{\mp 1}$, which agrees with the Ward identity: $\omega^2 G_+(\omega,k=0)=G_-(\omega,k=0)$ \cite{Kovtun:2005ev}. Eq.~\eqref{eq:mainS} therefore fixes the frequencies to which the poles converge as $k \to 0$. The same conclusion about the $k=0$ spectrum can be reached for the self-dual case of the Einstein-Maxwell-axion theory \cite{Andrade:2013gsa} (see below) with analytically known correlators for all $k$ derived in Ref.~\cite{Davison_2015}. In fact, these conclusions hold in any self-dual situation with a single gapless mode. 

\begin{figure}[t!]
    \centering
    \includegraphics[width=\linewidth]{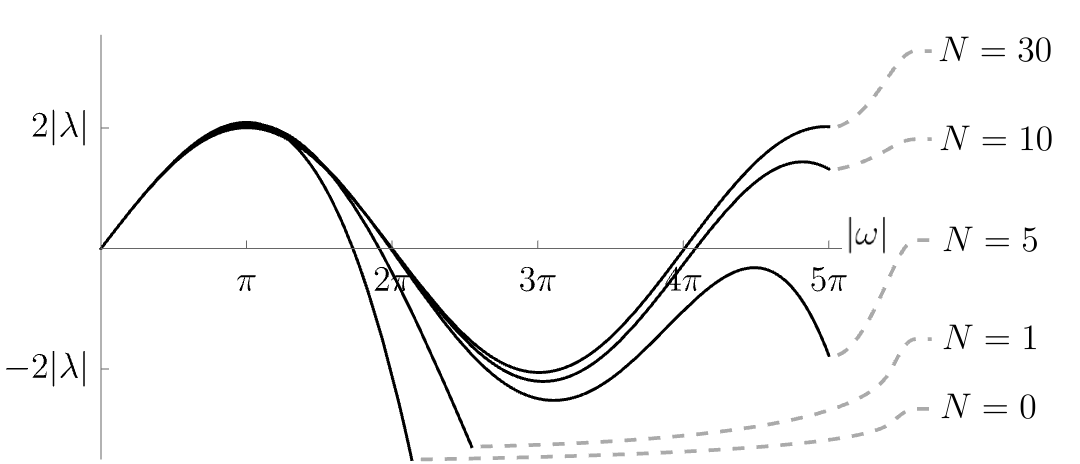}
    \caption{The plot of $S(i\abs{\omega})-S(-i\abs{\omega})$ for the spectra of $G_\pm^T$ with $k=1$ and $\beta=1$. $N$ corresponds to the number of gapped QNMs computed numerically \cite{Jansen:2017oag} and used in the truncation of the product $S$, so that the number of all modes is $4N+3$, combining gapped modes of both channels with the $3$ hydrodynamic modes. The value of $\lambda$ is $\lambda \approx -875$.}
    \label{fig:enter-label}
\end{figure}

Next, we consider an example of the duality relation \eqref{eq:duality}. In the AdS${}_4$-Schwarzschild black brane background, the correlators $G_\pm^T$ of $T^{\mu\nu}$ can be expressed as
\begin{align}
    G^T_+&=\frac{4}{\gamma k^4}\qty[\frac{\delta \sqrt{-g} \expval{T^{tt}}}{\delta g_{tt}}-2\bar\epsilon],\label{def:TT_plus} \\
    G^T_-&=\frac{4}{\gamma k^4}\qty[k^2\frac{\delta \sqrt{-g} \expval{T^{ty}}}{\delta g_{ty}}+\bar\epsilon\frac{6 \omega^2-k^2}{4}] ,\label{def:TT_minus}
\end{align}
where $\gamma$ is again some constant in terms of which the equilibrium energy density is given by $\bar\epsilon=\gamma(4 \pi/3\beta)^3$. Moreover, in thermal equilibrium, $\expval{T^{\mu\nu}}= \text{diag}(\bar\epsilon,\bar\epsilon/2,\bar\epsilon/2)$. In this case, $\omega_*$ is given by \cite{Grozdanov:2020koi,Grozdanov:2023txs,lenziDarbouxCovarianceHidden2021}
\begin{eqnarray}
    \omega_*=i \frac{\gamma k^4}{6 \bar\epsilon}. \label{def:asfreq}
\end{eqnarray}
The spectrum of $G^T_+$ contains a pair of gapless sound modes $\omega = \pm v_s k - i \Gamma k^2 + \mathcal{O}(k^3)$ and the spectrum of $G^T_-$ a diffusive mode $\omega = - i D k^2 + \mathcal{O}(k^4)$. The gapped modes are asymptotically described by the parameters $r \beta = 4\pi \sin (\pi/3)$ and $\theta = \pi/3$ \cite{Cardoso_2004,natario2005classification}, and, again, we have $\sigma=1$. The speed of sound is set by conformal symmetry to $v_s = 1/\sqrt{2}$. For small $k$, we get $\lambda \sim 2/i\beta \omega_*+\mathcal{O}(k^{-2})$. The spectral duality relation 
\eqref{eq:mainS} then alone (correctly) fixes the ratio between the momentum diffusivity $D$ and the sound attenuation $\Gamma$ to be $D/\Gamma = 2$ without any input from the effective theory of conformal hydrodynamics. Now, while Eq.~\eqref{eq:mainS} does not fix the spectrum at $k=0$ in a trivial way, it does, however, provide relations between the gapless dispersion relations and $\partial_k^n S(\omega,k=0)$. Furthermore, as in the self-dual case, it is easy to use Eq.~\eqref{eq:duality} along with the Ward identities \cite{Policastro:2002tn} to derive the $T=0$ correlators. Finally, we numerically demonstrate the validity of Eq.~\eqref{eq:mainS} in Fig.~\ref{fig:enter-label}.

In more general situations, explicit construction of the duality is a rather involved task. It can be done for example by finding the Darboux potentials of the relevant bulk wave equations (see Refs.~\cite{lenziDarbouxCovarianceHidden2021,Grozdanov:2023txs,glampedakisDarbouxTransformationBlack2017}). Instead, however, to use Eq.~\eqref{eq:mainS}, only the knowledge of $\omega_*$ is required, which can be computed in a simple manner from the Robinson-Trautman solution described in Ref.~\cite{Bakas_2014}. For details, see Appendix~\ref{appendix:recipe}.

Consider now the charged AdS${}_4$-Reissner-Nordstr\"{o}m black brane. A dual CFT in this state exhibits non-trivial couplings between $T^{\mu\nu}$ and $J^\mu$ correlators (see Ref.~\cite{Davison:2011uk}). However, by defining appropriate mixed operators, the system can be decoupled into two sectors \cite{Kodama:2003kk} for which we find that the corresponding algebraically special frequencies are
\begin{align}
    \omega_*&= i \frac{\gamma k^4}{6 \bar \epsilon}\left(\frac{1}{2}\pm \sqrt{\frac{1}{4}+\qty(\frac{2 Q \gamma}{3\overline{\epsilon}})^2 k^2}\right)^{-1}  ,\label{def:as_nordstrom}
\end{align}
where $Q$ is the charge. Hence, at finite $Q$, the large-$k$ asymptotics change from $k^4$ to $k^3$. Moreover, we see that as $Q\to 0$, one sector (with $\omega_*\to\infty$) reduces to the self-dual case of $G^J_\pm$ correlators and the other (with $\omega_*$ reducing to Eq.~\eqref{def:asfreq}) to the dual case of $G^T_\pm$ correlators discussed above. 

For our last example, we return to the uncharged Einstein-Maxwell-axion theory characterised by a translational-symmetry-breaking parameter $m$ \cite{Andrade:2013gsa}. Linearised bulk perturbations now decouple into three sectors: the gauge field perturbations, and two sectors that mix the axion and the metric perturbations, each with their even and odd channels. The gauge field correlators are self-dual for any value of $m$. The channel that reduces to purely metric perturbations at $m=0$ is controlled by 
\begin{equation}
    \omega_*=i \frac{\sqrt{2} k^2 (k^2+m^2)}{3 m_0 (m_0^2-m^2)}, \label{def:as_axion}
\end{equation}
where $\bar\epsilon=\gamma m_0(m_0^2-m^2)/2\sqrt{2}$ and $\beta=4\pi\sqrt{2} m_0 /(3m_0^2-m^2)$. At the self-dual point $m=m_0$, we see that, as required, $\overline{\epsilon}\rightarrow 0$ and $\omega_* \to \infty$. Moreover, in this limit, we can also explicitly compute \cite{Davison_2015}
\begin{equation} \label{eq:axion_S}
    S(\omega)=1+i\lambda \sinh{\frac{\beta \omega}{2}},
\end{equation}
where $\lambda$ depends on the considered correlator and can easily be found in closed form. Eq.~\eqref{eq:axion_S} explicitly verifies our spectral duality relation in Eq.~\eqref{eq:mainS}.  

Finally, we note that the algebraically special frequencies \eqref{def:as_nordstrom} and \eqref{def:as_axion} also provide new results for explicit lines of the algebraically special pole-skipping points at $\omega_*(k) = \pm \Omega_l$, and illustrate why no such lines are present at the self-dual points where they are `pushed to infinity'. This fact demonstrates another qualitative property distinguishing between the correlators in states that exhibit dual (cf.~Eq.~\eqref{eq:duality}) versus self-dual (cf.~Eq.~\eqref{eq:self_duality}) behaviour. 

\section{Discussion and future directions}

We presented new highly stringent constraints on the spectra of thermal holographic 3$d$ CFTs and, by nature of the AdS/CFT correspondence, on the QNMs of black holes in 4$d$ gravity (with matter fields). Our central result --- the spectral duality relation in Eq.~\eqref{eq:mainS} --- combined general properties of thermal correlators, the thermal product formula \cite{Dodelson:2023vrw} and dualities between 3$d$ CFTs correlators in different channels. Moreover, in the process of constructing these results, we also derived a new sum rule \eqref{eq:sumrule} on the thermal spectra, which holds in any dimension, irrespective of duality constraints. We expect our results to hold universally for broad classes of large-$N$ CFT and gravitational QNM spectra with applicability across subfields of physics. Furthermore, we note that the results of this paper can be readily generalised to spherical or hyperbolic boundary geometries.

While our work presented new infinite sets of constraints that could be used to make concrete predictions and verified in a number of holographic cases, further detailed studies into the mathematical structures underpinning the families of all possible spectra permitted by Eq.~\eqref{eq:mainS} are left to future work. In fact, such investigations should be crucial for finding the minimal set of necessary input needed to fix operator spectra uniquely, which was a question explored in Ref.~\cite{Grozdanov:2023tag}. Since our results were derived by using very general techniques, it will be particularly interesting to apply them to future investigations of CFTs without (known) holographic duals, e.g., to QED${}_3$ with a large number of flavours \cite{Giombi:2016fct,Romatschke:2019qbx} or other 3$d$ theories with known bosonic and fermionic particle-vortex dualities (see Refs.~\cite{sachdev_2011,Son_compositefermion,Karch:2016sxi,Seiberg:2016gmd,Son_selfdual,burgessParticleVortexDualityModular2001}). Other interesting questions that should be addressed through this new lens include a better understanding of the interplay between the hydrodynamic modes and the highly damped part of the spectrum using the operator product expansion techniques \cite{Caron-Huot:2009ypo,Dodelson:2023vrw} and the large-$k$ WKB methods \cite{Festuccia:2008zx,Fuini:2016qsc}. Finally, many future explorations of CFTs will surely investigate the analytic properties of thermal spectra at finite coupling and potential violations of duality constraints in higher-derivative theories  \cite{Myers:2010pk,Grozdanov:2016vgg}, as well as in theories without a large-$N$ limit. We believe that many such investigations may be dramatically simplified by considering the thermal generalisations of the S-duality (or particle-vortex duality) presented in this work.

\begin{acknowledgments}
We thank Guri Buza, Richard Davison, Giorgio Frangi, Pavel Kovtun, Ioannis Papadimitriou, Alexander Soloviev and Alexander Zhiboedov for valuable discussions on related topics. The work of S.G. was supported by the STFC Ernest Rutherford Fellowship ST/T00388X/1. The work is also supported by the research programme P1-0402 and the project N1-0245 of Slovenian Research Agency (ARIS). M.V. was supported by the STFC Studentship ST/X508366/1.
\end{acknowledgments}

\appendix
\section{A prescription for computing $\omega_*$}
\label{appendix:recipe}
We consider a family of black brane backgrounds
\begin{align}
        ds^2=-f(r)dt^2+\frac{dr^2}{f(r)}+r^2(dy^2 + dz^2) \label{def:metric}
\end{align}
linearly perturbed with a specific ansatz for the Robinson-Trautman solution \cite{Bakas_2014}:  
\begin{subequations}
    \label{eq:perturbations}
\begin{align}
    \delta g_{tt}/f(r)&=f(r) \delta g_{rr}=\delta g_{tr}=\xi(r) e^{-i \omega t+i k z},\\
    \delta g_{yy}&=\delta g_{zz}=r^2 \zeta e^{-i\omega (t+r_*) + i k z}, \\
    \delta g_{ty}&=\delta g_{tz}=\delta g_{ry}=\delta g_{rz}=\delta g_{yz}=0,
\end{align}
\end{subequations}
for some $\xi(r)$ and $\zeta$, and with the tortoise coordinate $r_*$ defined by $dr_*/dr=1/f(r)$. All the other fields in the theory are perturbed with full generality. The ansatz only solves the linearised equations of motion for the algebraically special frequencies $\omega=\omega_*$. Since the perturbation \eqref{eq:perturbations} is in the longitudinal channel, and ingoing, the ansatz gives the sign of $\omega_*$ so that $G_+(\omega_*)=0$. While it is not clear whether this always gives the correct $\omega_*$, or that such a solution implies the existence of duality relations, it works in all the considered cases and we expect it to hold on very general grounds.

\section{Derivation of the sum rule}\label{app:sum}

To derive the sum rule in Eq.~\eqref{eq:sumrule}, we consider a class of retarded correlators $G(\omega)$ with the following asymptotic property:
\begin{equation}
\lim_{\omega\rightarrow\infty}\partial_\omega\qty[G(\omega)-G(-\omega)]=0. \label{eq:Gconvergence}
\end{equation}
For example, this condition is satisfied for all $G(\omega)$ with $m_\pm \leq 0$ (cf.~Eq.~\eqref{eq:Gasy}). Such cases include the $G^J_+$ (cf.~Eq.~\eqref{def:JJ}) and the $G^T_+$ (cf.~Eq.~\eqref{def:TT_plus}) correlators in an uncharged thermal state.
Such a correlator can be conveniently decomposed into two functions:
\begin{equation}
G(\omega)=\tilde{G}(\omega)+c(\omega),
\end{equation}
where $c(\omega)=c(-\omega)$ and $\tilde G(\omega)$ with the following asymptotics:
\begin{equation}
    \lim_{\omega\rightarrow\infty} \partial_\omega \tilde{G}(\omega)=0.
\end{equation}
Here, $c(\omega)$ is a real counterterm with no poles that can also vanish identically. Furthermore, we assume that it is sufficiently smooth around $\omega=0$ so that 
$\partial_\omega c(\omega)|_{\omega=0}=0$. Analiticity of $\partial_\omega G(\omega)$ in  the upper complex $\omega$ half-plane then allows us to write (see Refs.~\cite{Gulotta_2011,Romatschke_2009})
\begin{eqnarray}
    \partial_\omega \tilde{G}(\omega)=\int_\mathbb{R}\frac{d\omega'}{2\pi i}\frac{G(\omega')-G(-\omega')}{(\omega'-\omega-i0^+)^2},\label{eq:KK}
\end{eqnarray}
where the convergence of the integral is equivalent to the assumption \eqref{eq:Gconvergence}. The sum rule \eqref{eq:sumrule} then follows once we evaluate Eq.~\eqref{eq:KK} at $\omega=0$, use the thermal product formula \eqref{eq:TPF} and the definition of $\lambda$ in Eq.~\eqref{def:lambda}. Generalisations of the sum rule \eqref{eq:sumrule} to situations in which the assumption \eqref{eq:Gconvergence} does not hold should be possible with the inclusion of appropriate additional counterterms into the integral \eqref{eq:KK}.

\bibliography{milebib}

\end{document}